\documentclass[iop]{emulateapj}
\usepackage{apjfonts}
\usepackage{hyperref}
\usepackage{graphicx}
\usepackage{bigstrut}

\newcommand{\exo}{EXO 0748--676}
\newcommand{\ter}{1E 1724--3045}
\newcommand{\xte}{{\it RXTE}}

\newcommand{\sa}{$S_{\rm a}$}
\newcommand{\lb}{$L_{\rm b}$}
\newcommand{\lh}{$L_{\rm h}$}

\shorttitle{Geometric Nature of Low-Frequency QPOs}
\shortauthors{Homan et al.}

\begin{document}

\title{On the geometric nature of  low-frequency quasi-periodic oscillations in neutron-star low-mass X-ray binaries}

\author{Jeroen Homan\altaffilmark{1, 2},   Joel K.\ Fridriksson\altaffilmark{3}, Ronald A.\ Remillard\altaffilmark{1}}

\altaffiltext{1}{MIT Kavli Institute for Astrophysics and Space Research, 77 Massachusetts Avenue 37-582D, Cambridge, MA 02139, USA; jeroen@space.mit.edu}

\altaffiltext{2}{SRON, Netherlands Institute for Space Research, Sorbonnelaan 2, 3584 CA Utrecht, The Netherlands}

\altaffiltext{3}{ Anton Pannekoek Institute for Astronomy, University of Amsterdam, Postbus 94249, 1090 GE Amsterdam, The Netherlands}

\begin{abstract}

We report on a detailed analysis of the so-called $\sim$1 Hz quasi-periodic oscillation (QPO) in the eclipsing and dipping neutron-star low-mass X-ray binary \exo. This type of QPO has previously been shown to have a geometric origin. Our study focuses on the evolution of the QPO  as the source moves through the color--color diagram, in which it traces out an atoll-source-like track. The QPO frequency increases from $\sim$0.4 Hz in the hard state to $\sim$25 Hz as the source approaches the soft state. Combining power spectra based on QPO frequency reveals additional features that strongly resemble those seen in non-dipping/eclipsing atoll sources. We show that the low-frequency QPOs in atoll sources and the $\sim$1 Hz QPO in \exo\ follow similar relations with respect to the noise components in their power spectra. We conclude that the frequencies of both types of QPOs are likely set by (the same) precession of a misaligned inner accretion disk. For high-inclination systems, like \exo, this results in modulations of the neutron-star emission due to obscuration or scattering, while for lower-inclination systems the modulations likely arise from relativistic Doppler boosting and light-bending effects. 

\end{abstract}

\keywords{accretion, accretion disks -- stars: neutron -- X-rays: binaries -- X-rays: individual (\exo)}

\section{Introduction}\label{sec:intro}

Low-frequency quasi-periodic oscillations (LF-QPOs) are a common phenomenon in the X-ray light curves of low-mass X-ray binaries. In power spectra they show up as narrow peaks with frequencies ranging from $\sim$1 to $\sim$70 Hz in neutron-star low-mass  X-ray binaries (NS-LMXBs) and from $\sim$0.01 to $\sim$30 Hz in black-hole low-mass X-ray binaries (BH-LMXBs). Comparisons of the power spectra of NS-LMXBs and BH-LMXBs suggest that the most common types of LF-QPOs in these systems (as well as some of the other variability components) may have a similar origin \citep{wiva1999,psbeva1999,bepsva2002}

For the most common type of LF-QPOs in BH-LMXBs, the so-called type-C QPOs \citep{cabest2005}, there is increasing evidence that points toward a geometric origin: these QPOs are stronger in sources that are viewed at higher inclinations \citep{schomi2006,heutkl2014,mocahe2015}, and the strength of the  Fe$\alpha$ emission line in GRS 1915+105 has been found to be strongly modulated as part of the type-C QPO cycle \citep{miho2005,inva2015}. Both these results are consistent with the QPO being generated by a misaligned and precessing inner accretion disk \citep{schomi2006}, with relativistic Lense--Thirring precession being a commonly suggested mechanism \citep[e.g.,][]{indofr2009}.

In seven dipping and/or eclipsing NS-LMXBs a type of LF-QPO has been observed whose properties suggest (even more directly) that it also has a geometric origin \citep{ho2012}. These QPOs are often referred to as the ``$\sim$1 Hz QPOs'', due to the fact that they were initially discovered around 1 Hz in three different sources \citep{hojowi1999,jovawi1999,jovaho2000}. The geometric nature of these QPOs was deduced from the fact that these QPOs persist during type I (thermonuclear) X-ray bursts with a fractional amplitude that is consistent with that observed during the persistent  (non-burst) emission, and that their strength is nearly independent of energy. Both these properties suggests that (part of) the emission from the neutron star is obscured or scattered out of the line of sight by the accretion flow. \citet{ho2012} recently suggested that an inclined and precessing inner accretion flow might be a possible explanation for the observed properties of the $\sim$1 Hz QPOs. The misalignment angle between the inner flow and binary plane would have to be $\sim$$15^\circ$--$20^\circ$ for this type of QPO to be observed only in dipping and/or eclipsing NS-LMXBs. The relation between these $\sim$1 Hz QPOs and the LF-QPOs observed in the non-dipping/eclipsing NS-LMXB is not clear at this point. Although \citet{ho2012} pointed out some similarities (e.g., both being mainly observed in the harder spectral states) the $\sim$1 Hz QPOs in the dipping/eclipsing sources have not been studied in enough detail to make a proper comparison. 

In this paper we investigate the $\sim$1 Hz QPO observed in the dipping and eclipsing NS-LMXB \exo. Of the seven sources in which $\sim$1 Hz QPOs have been observed, \exo\ has the largest reported frequency range. It also has the highest number of {\it Rossi X-ray Timing Explorer} (\xte) observations and is known to make regular state transitions. We focus on how the $\sim$1 Hz QPO evolves along the tracks traced out in the color--color diagram (CD; Section \ref{sec:exo}) and perform a comparison of the observed power spectra with those of non-dipping/eclipsing NS-LMXBs (Section \ref{sec:atoll}). Our results are discussed in Section \ref{sec:disc}.

\begin{figure}[t] 
\centerline{\includegraphics[width=8.5cm]{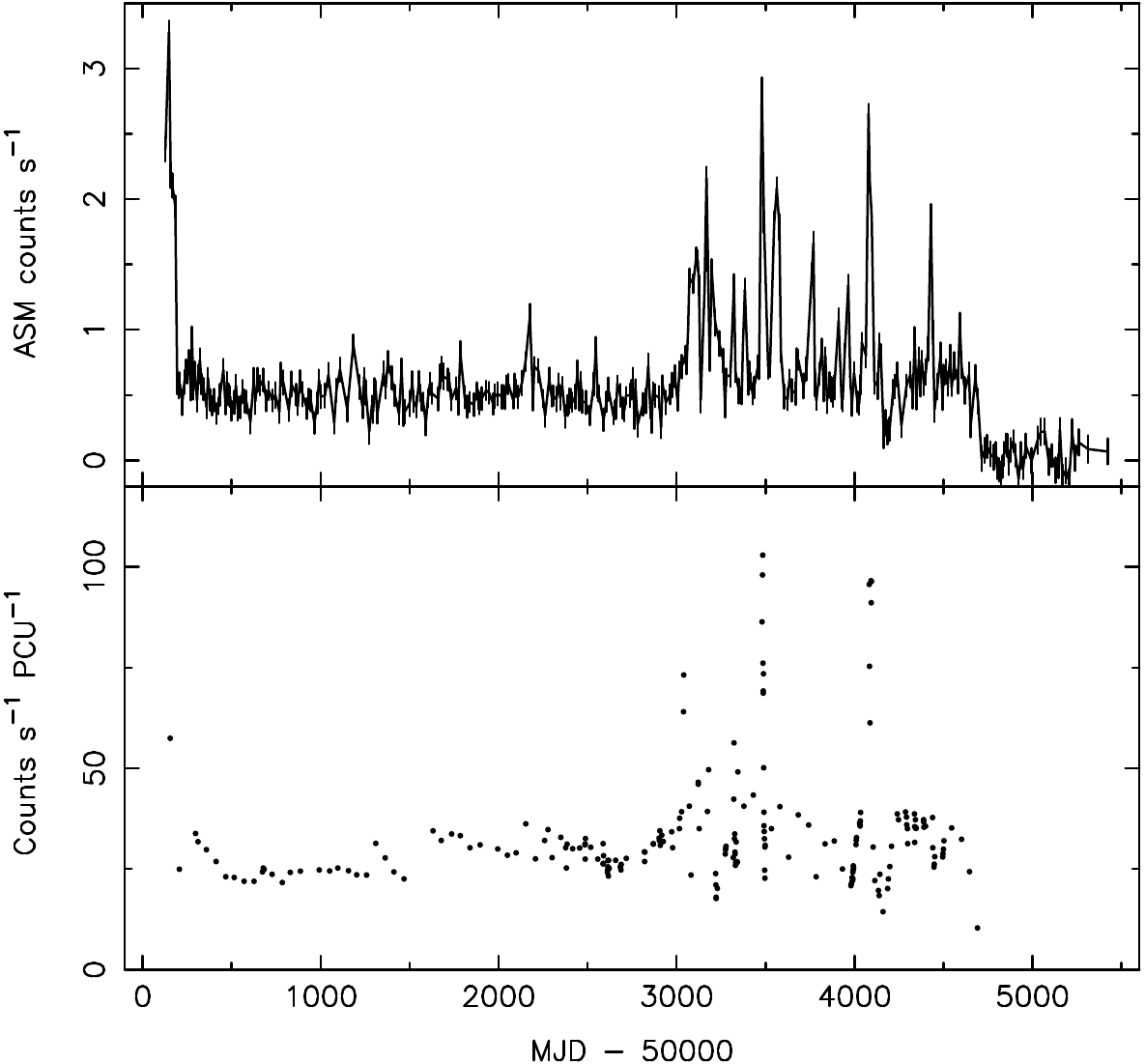}}
\caption{Long-term light curves of \exo. In the top panel we show 7-day averages from the {\it RXTE}/ASM ($\sim$1.5--12 keV). In the bottom panel we show data from the {\it RXTE}/PCA ($\sim$2--60 keV), with each data point representing the average count rate of a single group. Errors on the PCA count rates are smaller than the size of the plot symbols. Around MJD 54700 (2008 August) the source returned to quiescence.} 
\label{fig:curve}
\vspace{0.2cm}
\end{figure}

\section{Observations and data reduction}\label{sec:obs}

The observations of \exo\ analyzed in this paper were made with the Proportional Counter Array \citep[PCA;][]{jamara2006} onboard \xte\ \citep{brrosw1993}. \exo\ was observed 752 times with the PCA during the lifetime of \xte. A number of observations had to be discarded for various reasons: they were made after the source had returned to quiescence \citep{worawo2008a,worawo2008b}, important data modes were missing, or not enough data were left to produce power spectra of the required length (128 s) after the recommended filtering criteria had been applied. This left 718 useful observations. To increase the sensitivity to QPOs, observations were combined into groups. Consecutive observations were added together if the time gap between them was less than 12 hr, resulting in 203 groups of observations. In all almost all cases these groups covered periods less than a day, but for seven groups the period was longer (up to $\sim$2 days). These seven groups were split into two or three shorter segments, covering less than a day. The final number of groups was 211, with individual groups containing between 1 and 11 observations. The group exposure times ranged from $\sim$350 s to $\sim$40 ks, with an average of $\sim$8 ks. Note that this process of combining observations is justified by the fact that the $\sim$1 Hz QPO in \exo\ was seen to vary in frequency by less than $\sim$15\% on a timescale of a day \citep{hojowi1999}.

We used {\tt standard 2} mode data to create light curves, hardness--intensity diagrams (HIDs), and CDs. Data from all available Proportional Counter Units (PCUs) were used.  Following the procedures laid out in \citet{frhore2015},  count rates from each PCU were corrected for changes in the response, using long-term light curves of the Crab, and then normalized to those of PCU 2. We define a soft color as the net counts in the 4.0--7.3 keV band divided by those in the 2.4--4.0 keV band, and a hard color (hardness) as the net counts in the 9.8--18.2 keV band divided by those in the 7.3--9.8 keV band. The intensity we use for the HIDs is the net count rate per PCU in the 2--60 keV band. The {\tt standard 2} data were barycenter corrected and eclipses were removed using the ephemeris from \citet{worawo2009}. Type I X-ray bursts were removed using {\tt standard 1} mode data, following the procedures laid out in \citet{relico2006}. The {\tt standard 2} and high-time-resolution data were filtered following a similar procedure. Absorption dips were removed from the {\tt standard 2} data, since they strongly affect the average count rates and colors of the observations. The removal of dips had to be done manually and was based on visual inspection of the light curves and soft-color curves. Given the strong differences between observations in terms of dipping activity it was difficult to remove dips in a fully consistent manner for the entire data set.  We did not remove the dips from the high-time-resolution data used for the creation of the power spectra, since the $\sim$1 Hz QPO can still be detected during dips \citep{hojowi1999}.

Power spectra were created from light curves extracted from the high-time-resolution data modes. Light curves were created with a time resolution of 1/8192 s (except for the first observation, 10068-03-01-00, which had a maximum time resolution of 1/2048 s), covering absolute PCA channels 0--35. This channel band effectively corresponds to an energy range of $\sim$1.5--9.5 keV during Gain Epoch 1, increasing to $\sim$2.1--15 keV during Gain Epochs 4 and 5. We note that more than 600 of our observations were made during Gain Epochs 4 and 5. To create the power spectra we performed Fast Fourier Transformations of 128 s light-curve segments. The power spectra were averaged for each group and rms normalized \citep{beha1990,mikiki1991,va1995b}.  These
normalized power spectra were rebinned logarithmically to 70 bins per frequency decade; for plotting purposes the power spectra were sometimes rebinned even further.  Prior to fitting we subtracted the deadtime-modified Poisson level, which we approximated by taking the average power in an interval of a few 100 Hz centered around $\sim$2000\,\,Hz. While this method may introduce some uncertainties in the properties of the high-frequency features, it should have a negligible effect on the low-frequency features in the power spectra (which are the focus of this paper). The resulting power spectra were fitted up to 2000 Hz with a combination of a power law
($P(\nu)\propto\nu^{-\alpha}$),  and one or more
Lorentzians ($P(\nu)=(r^2\Delta/\pi)[\Delta^2 +
(\nu-\nu_0)^2]^{-1}$). Here $\nu_0$ is the centroid
frequency, $\Delta$ the half-width-at-half-maximum, and $r$ the
integrated fractional rms (from $-\infty$ to $\infty$). Instead of
$\nu_0$ and $\Delta$ we will quote the frequency $\nu_{\rm max}$ at which
the Lorentzian attains its maximum in $\nu P(\nu)$ and the quality
factor, Q, where $\nu_{\rm max}=\nu_0 (1 + 1/4Q^2)^{1/2}$ and
$Q=\nu_0/2\Delta$ \citep{bepsva2002}. The fractional rms amplitudes
quoted in this paper were the integrated power between 0 and $\infty$ Hz
for the Lorentzians, and between 0.01 and 100 Hz for the power-law
component. Errors on fit parameters were determined using
$\Delta\chi^2=1$ (i.e. 68\% confidence).

\begin{figure}[t] 
\centerline{\includegraphics[width=8.5cm]{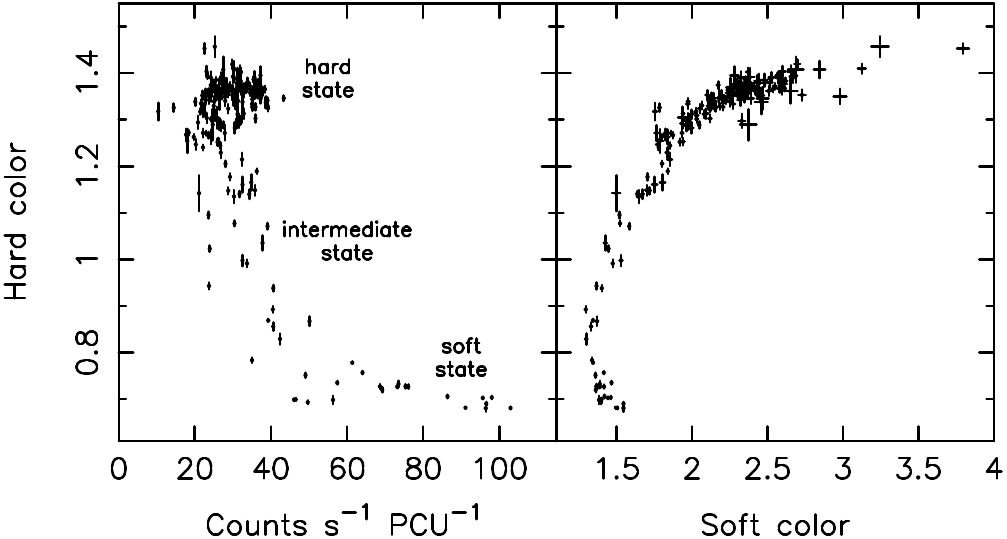}}
\caption{Hardness--intensity diagram (left) and color--color diagram (right) of \exo. Each data point represents the average of a single group. See Section \ref{sec:obs} for definition of colors.} 
\label{fig:cdhid}
\vspace{0.2cm}
\end{figure}

\begin{figure}[t] 
\centerline{\includegraphics[width=7.5cm]{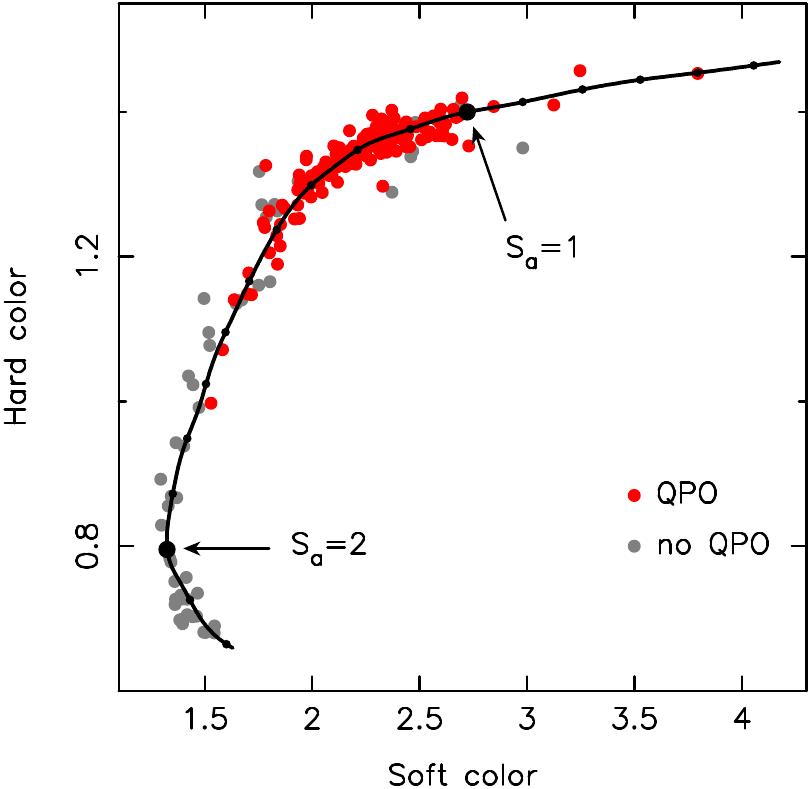}}
\caption{Color--color diagram of \exo, showing groups with $\sim$1 Hz QPO detections (red) and those without (gray). The black line shows the spline used to calculate the $S_{\rm a}$ values for each data point. The two large black circles indicate the locations of $S_{\rm a} = 1$ and $S_{\rm a} = 2$. The smaller black circles represent steps in  $S_{\rm a}$ of 0.1. Error bars are omitted for clarity.} 
\label{fig:cd}
\end{figure}

\newpage

\section{Analysis \& results}\label{sec:results}

\subsection{The $\sim$1 Hz QPO in \exo}\label{sec:exo}

In Figure \ref{fig:curve} we show  long-term  light curves of \exo\ from the \xte\ All-Sky Monitor\footnote{Public ASM data were obtained from \url{http://xte.mit.edu}.} \citep[ASM;][]{lebrcu1996} and the PCA.  The ASM light curve shows 7-day averages, while the PCA light curve  shows the average count rate for each group.  Especially during the second half of  \xte's lifetime \exo\ was quite variable, with occasional several-week-long outbursts. In Figure \ref{fig:cdhid} we show the HID  and CD corresponding to the PCA light curve. The patterns traced out in those diagrams are very similar to those of some of the low-luminosity (atoll) NS-LMXBs (see, e.g., \citet{gldogi2007} and \citet{fr2011} for collections of atoll-source CDs/HIDs). As already pointed out by \citet{pomufe2014} and \citet{mufemo2014},  hard and soft states can be identified in the CD/HID of \exo. In the HID in Figure \ref{fig:cdhid} we label the hard, intermediate, and soft states; these states are often also referred to as the extreme island state, the island state and the (lower) banana state, respectively. A comparison between the CD of \exo\ and those of the atoll sources presented in \citet{fr2011} indicates that the soft color of \exo\ spans an unusually large range. This is likely the result of incomplete dip removal, which mostly affects that low energy bands. We note that before the dip removal the soft color range was $\sim$1.3--5.1.

Based on a visual inspection of all 211 power spectra for the presence of a $\sim$1 Hz QPO, we selected 171 for further analysis. Following the analysis of \citet{hojowi1999} these 171 power spectra were initially fit with a combination of a power law and a single Lorentzian. While in most cases this model resulted in acceptable fits (average $\chi^2_{\rm red}\approx1.18$), in some power spectra the QPO profile was not accurately fit because of a broad underlying noise component. Adding a zero-centered Lorentzian to the fit function solved this issue and resulted in an average $\chi^2_{\rm red}$ of $\sim$1.05.  QPOs with single-trial significances\footnote{The integrated power divided by its negative error.} larger than 3$\sigma$ were detected  in the power spectra of 157 groups, with frequencies ranging from  $\sim$0.43 to $\sim$25.5 Hz. Together these 157 groups represent 558 ($\sim$78\%) of the 718 observations we analyzed.

\begin{figure}[t] 
\centerline{\includegraphics[width=8.cm]{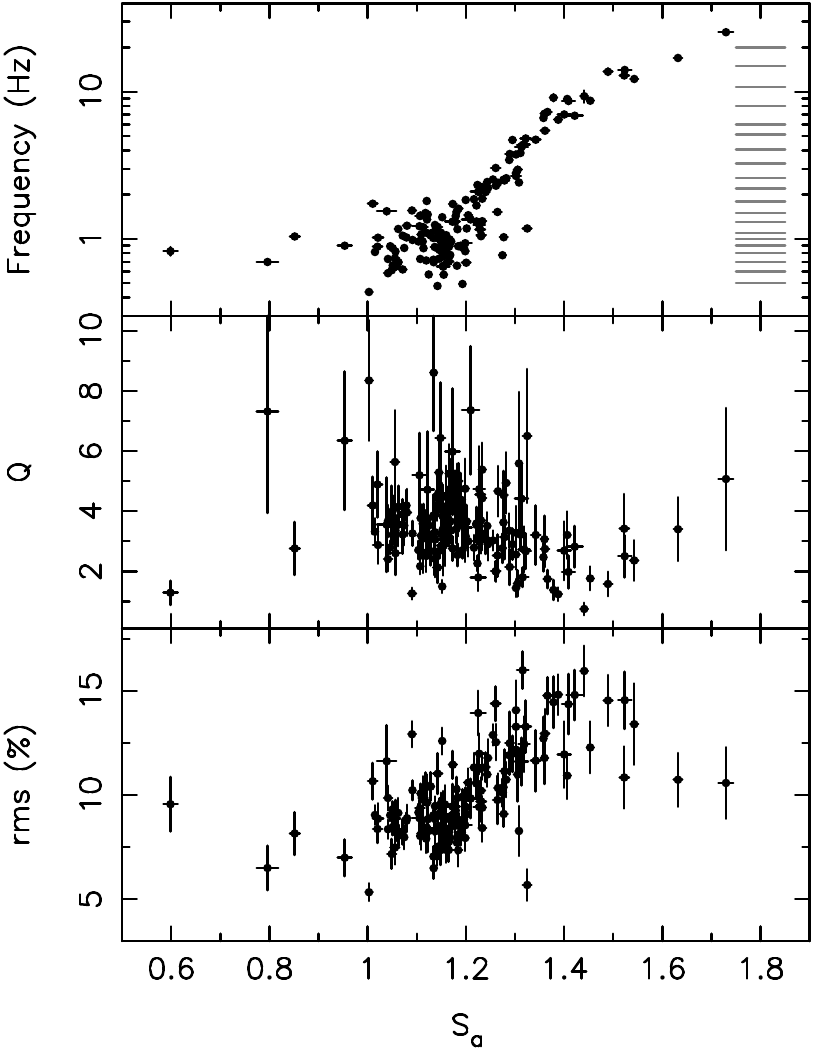}}
\caption{Properties of the $\sim$1 Hz QPO in \exo\ as a function of $S_{\rm a}$. From top to bottom we show frequency, $Q$-value, and fractional rms. The errors on the frequency are typically smaller than the symbol size. The 20 gray horizontal bars in the upper panel show the boundaries between intervals used for grouping power spectra based on QPO frequency (see main text). } 
\label{fig:sz}
\end{figure}

To see in which states the $\sim$1 Hz QPO was detected we show a color-coded CD in Figure \ref{fig:cd}; the groups with LF-QPO detections are colored red, and those without are colored gray. It is clear from this figure that the $\sim$1 Hz QPO was predominantly detected in the hard state, with a few detections in the intermediate state. In order to study how the QPO properties evolve along the track in the CD, we parameterized the position along the track by manually drawing a spline onto which all data points were projected \citep[see, e.g.,][]{diva2000}.  To correct for the differences in scale between the horizontal and vertical axes in Figure \ref{fig:cd}, the soft-color values were scaled down by a factor of 0.286 during the projection procedure, thus ensuring that the data points were projected onto the nearest spline point in the figure. Two normal points,  depicted by the large black circles in Figure \ref{fig:cd}, were chosen to set the scale for the track parameter \sa. $S_{\rm a}=1$ was chosen to fall close to the end of the dense cluster of points in the hard state of \exo, while $S_{\rm a}=2$ was chosen to coincide with the boundary of the intermediate state and the soft state (most clearly seen in the HID in Figure \ref{fig:cdhid}; hard color $\sim$0.8). The smaller black circles represent steps of 0.1 in \sa. Using this parametrization we find that no $\sim$1 Hz QPO was detected significantly for \sa\ values greater than 1.73. Note that the observation that showed the 695 Hz QPO reported by \citet{hova2000} has an \sa\ value of $\sim$2.1.

In Figure \ref{fig:sz} we show the QPO frequency, $Q$-value and rms amplitude as a function of \sa. The QPO frequency shows a clear increase with \sa, although there is considerable scatter at low \sa\ values. The $Q$-value shows no clear dependence on \sa, while the rms appears to increase toward higher \sa.

\begin{figure}[t] 
\centerline{\includegraphics[width=8.cm]{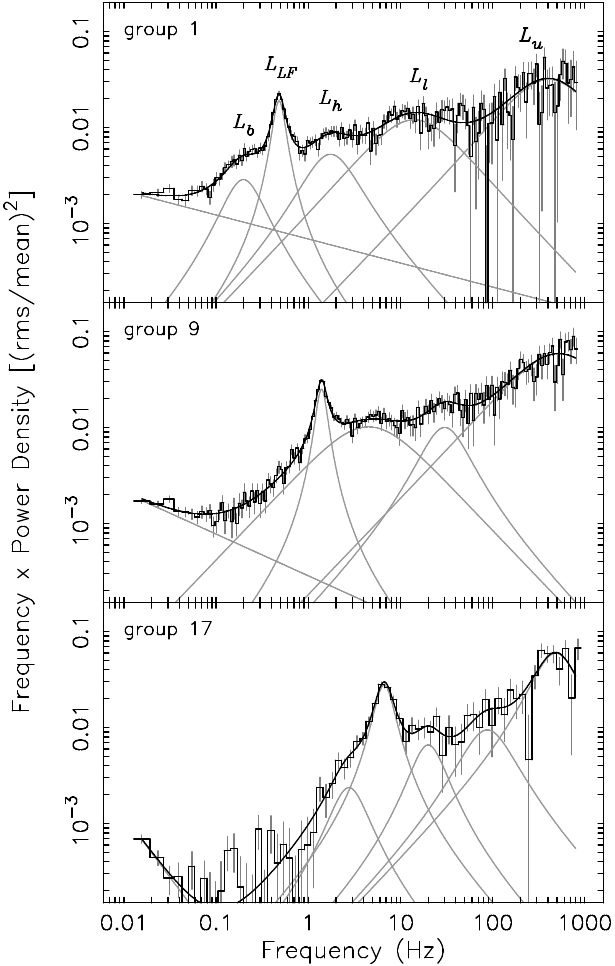}}
\caption{Three representative power spectra of \exo\ from frequency-selected groups. Fits to the power spectra are shown as solid black curves, while individual components are shown in gray. The various power-spectral components are identified in the top panel. The $\sim$1 Hz QPO is marked as $L_{\rm LF}$.} 
\label{fig:pds}
\vspace{0.2cm}
\end{figure}

Part of the scatter observed in Figure \ref{fig:sz} is likely caused by the fact that we were not able to perfectly remove all dipping intervals from the data. Since dipping is most prevalent in the hard state (i.e., at low \sa) and increases the soft color, the removal of dips will move observations parallel to the spline in the CD, towards higher \sa. Various gradations of incomplete dip removal will therefore result in a range of \sa\ values for observations whose non-dipping emission would have the same \sa\ value. To reduce the impact from these effects, and also to boost the signal-to-noise ratio of our power spectra, we performed additional grouping of our observations. Observations from groups with similar QPO frequencies were combined into 21 new groups. At the low-frequency end the width of the frequency selections for the new groups was 0.1 Hz, increasing to larger values for higher frequencies; the gray horizontal lines in the top panel of Figure \ref{fig:sz} indicate the boundaries separating the 21 frequency-based groups. These new groups contained between 1 and 17 of the old groups, and between 1 and 69 of the original observations.   To determine \sa\ values for each of the new groups we reprojected the averaged CD data points onto the previously used spline. 

\begin{figure}[t] 
\centerline{\includegraphics[width=8.cm]{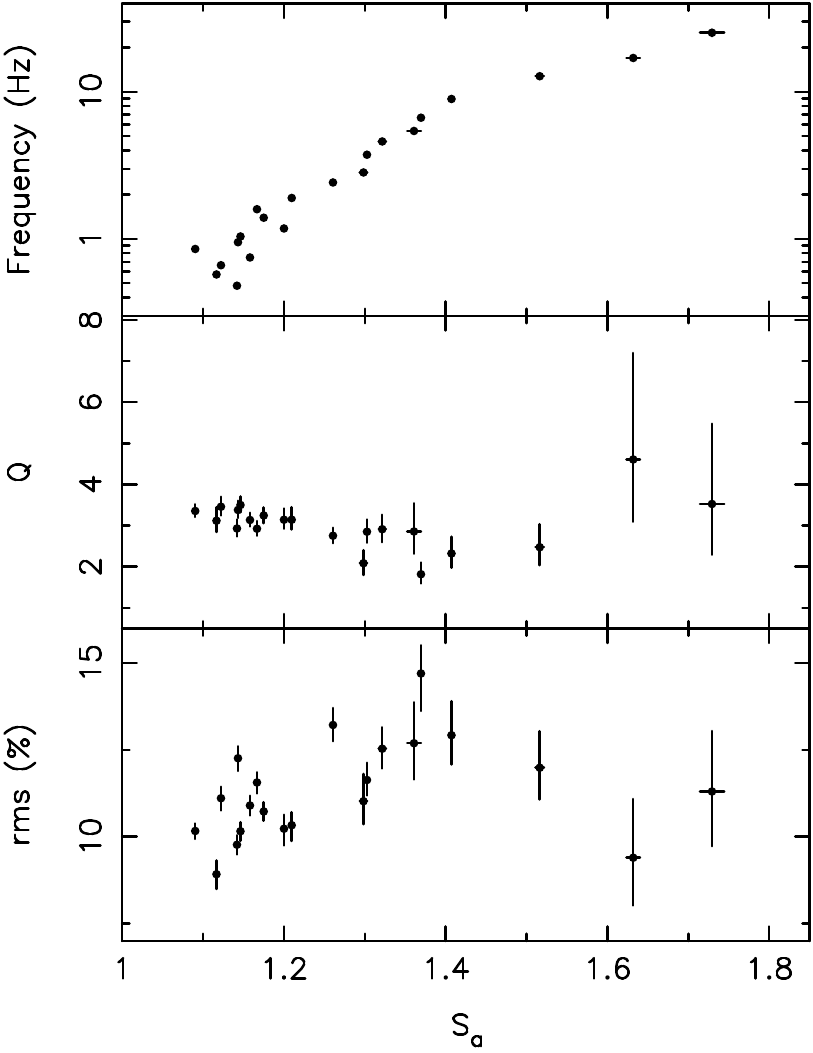}}
\caption{Properties of the $\sim$1 Hz QPO in \exo\ as a function of $S_{\rm a}$. From top to bottom we show frequency, $Q$-value, and fractional rms. The errors on the frequency are typically smaller than the symbol size.} 
\label{fig:sz_meta}
\end{figure}

\begin{center}
\begin{deluxetable*}{ccccccccccccc}
\tabletypesize{\scriptsize}
\tablewidth{0pt}
\tablecaption{Power Spectral Fit Parameters\label{tab:pds}} 
\tablehead{
\colhead{} & \colhead{} &
\multicolumn{3}{c}{L$_{\rm LF}$} & \colhead{} &
\multicolumn{3}{c}{L$_{\rm b}$} & \colhead{} &
\multicolumn{3}{c}{L$_{\rm h}$} \bigstrut[b] \\
\cline{3-5} \cline{7-9} \cline{11-13}
\colhead{} & \colhead{} &
\colhead{$\nu_{\rm max}$} & \colhead{$Q$} &  \colhead{rms} & \colhead{} &
\colhead{$\nu_{\rm max}$} & \colhead{$Q$} &  \colhead{rms} & \colhead{} &
\colhead{$\nu_{\rm max}$} & \colhead{$Q$} &  \colhead{rms} \bigstrut[t] \\
\colhead{No.$^a$} & \colhead{\sa} &
\colhead{(Hz)} & \colhead{} &  \colhead{(\%)} & \colhead{} &
\colhead{(Hz)} & \colhead{} &  \colhead{(\%)} & \colhead{} &
\colhead{(Hz)} & \colhead{} &  \colhead{(\%)}  
}
\startdata
1 	& 1.142$\pm$0.003	& 0.483$\pm$0.003	& 2.9$\pm$0.2			& 9.8$\pm$0.3 	& & 0.20$\pm$0.11			& 0.83$\pm$0.19	\bigstrut[tb]			& 6.5$\pm$0.7			& & 1.75$\pm$0.13		& 0.58$\pm$0.19			& 9.8$\pm$1.3 			\\
2 	& 1.116$\pm$0.003	& 0.574$\pm$0.004	& 3.1$\pm$0.3			& 8.9$\pm$0.4	& & 0.19$\pm$0.16			& 0.7$\pm$0.3		\bigstrut[tb]		& 5.9$\pm$0.9			& & 2.1$\pm$0.3			& 0.2$\pm$0.2			& 12.9$^{+2.4}_{-1.5}$ 	\\   
3 	& 1.122$\pm$0.003	& 0.665$\pm$0.003	& 3.5$\pm$0.2			& 11.1$\pm$0.3	& & 0.27$\pm$0.19			& 0.8$\pm$0.2	\bigstrut[tb]			& 6.7$\pm$0.8			& & 2.11$\pm$0.16		& 0.34$\pm$0.15			& 14.8$\pm$1.3 			\\   
4 	& 1.158$\pm$0.001	& 0.749$\pm$0.003	& 3.14$\pm$0.17			& 10.9$\pm$0.3	& & 0.27$\pm$0.16			& 0.8$\pm$0.2	\bigstrut[tb]			& 5.1$\pm$0.6			& & 2.25$\pm$0.10		& 0.42$\pm$0.09			& 15.8$\pm$1.1 			\\  
5 	& 1.090$\pm$0.002	& 0.857$\pm$0.003	& 3.36$\pm$0.16			& 10.2$\pm$0.2	& & 0.31$\pm$0.15			& 1.7$\pm$0.4	 \bigstrut[tb]			& 3.5$\pm$0.4			& & 2.66$\pm$0.18		& 0.23$\pm$0.07			& 16.8$\pm$1.0 			\\   
6 	& 1.143$\pm$0.001	& 0.952$\pm$0.004	& 3.4$\pm$0.2			& 12.3$\pm$0.4	& & 0.3$\pm$0.2				& 1.4$^{+0.7}_{-0.4}$ \bigstrut[tb]		& 4.0$\pm$0.7			& & 2.8$\pm$0.2			& 0.30$\pm$0.08			& 18.2$\pm$1.2 			\\  
7 	& 1.146$\pm$0.002	& 1.040$\pm$0.004	& 3.5$\pm$0.19			& 10.2$\pm$0.3	& & 0.40$\pm$0.03			& 1.4$^{+0.7}_{-0.4}$ \bigstrut[tb]		& 3.3$\pm$0.7			& & 3.2$\pm$0.3			& 0.23$\pm$0.10			& 15.7$\pm$1.7			\\  
8 	& 1.200$\pm$0.002	& 1.182$\pm$0.007	& 3.1$\pm$0.2			& 10.2$\pm$0.4	& & 0.50$\pm$0.05				& 0.9$^{+0.9}_{-0.5}$ \bigstrut[tb]		& 4.3$\pm$1.3			& & 3.4$\pm$0.3			& 0.38$\pm$0.19			& 13.9$^{+2.0}_{-1.3}$	\\  
9 	& 1.175$\pm$0.002	& 1.397$\pm$0.007	& 3.25$\pm$0.19			& 10.7$\pm$0.3	& & \nodata 				& \nodata		\bigstrut[tb]				& \nodata				& & 4.5$\pm$0.5			& 0.04$\pm$0.08			& 17.6$\pm$0.9			\\  
10 	& 1.167$\pm$0.002	& 1.598$\pm$0.009	& 2.92$\pm$0.18			& 11.6$\pm$0.3	& & \nodata 				& \nodata		\bigstrut[tb]				& \nodata				& & 5.7$\pm$0.7 		& 0$^b$ 				& 18.3$\pm$0.8			\\  
11 	& 1.209$\pm$0.002	& 1.901$\pm$0.013	& 3.1$\pm$0.3			& 10.3$\pm$0.4	& & \nodata 				& \nodata		\bigstrut[tb]				& \nodata				& & 4.5$\pm$1.2 		& 0.09$\pm$0.12			& 15.3$^{+1.6}_{-0.5}$			\\  
12 	& 1.261$\pm$0.002	& 2.430$\pm$0.014	& 2.76$\pm$0.19			& 13.2$\pm$0.5	& & 1.09$^{+0.47}_{-0.18}$	& 0.6$\pm$0.3		\bigstrut[tb]			& 5.7$\pm$1.9			& & 7.4$\pm$0.7			& 0.3$\pm$0.2			& 15.4$\pm$1.4  			\\  
13 	& 1.298$\pm$0.005	& 2.84$\pm$0.05		& 2.1$\pm$0.3			& 11.0$\pm$0.7	& & \nodata 				& \nodata		\bigstrut[tb]				& \nodata				& & 7.1$^{+1.8}_{-1.2}$	& 0$^b$					& 13.3$\pm$1.0			\\  
14 	& 1.303$\pm$0.003	& 3.75$\pm$0.03		& 2.9$\pm$0.3			& 11.6$\pm$0.5	& & \nodata 				& \nodata 	\bigstrut[tb]					& \nodata				& & 8.0$\pm$1.0			& 0$^b$ 				& 14.9$\pm$0.9 			\\  
15 	& 1.322$\pm$0.005	& 4.61$\pm$0.05		& 2.9$\pm$0.3			& 12.5$\pm$0.6	& & \nodata 				& \nodata		\bigstrut[tb]				& \nodata				& & 7.6$\pm$1.2 		& 0$^b$					& 15.3$\pm$1.0  			\\  
16 	& 1.361$\pm$0.009	& 5.43$\pm$0.11		& 2.9$\pm$0.6			& 12.7$\pm$1.1	& & \nodata 				& \nodata		\bigstrut[tb]				& \nodata				& & 15$^{+8}_{-5}$ 		& 0$^b$					& 13.1$^{+2.5}_{-1.0}$  			\\  
17 	& 1.370$\pm$0.003	& 6.68$\pm$0.10		& 1.8$\pm$0.3			& 14.7$\pm$0.9	& & 2.8$\pm$0.9 			& 1.0$^{+0.8}_{-0.4}$ 	\bigstrut[tb]		& 5.6$^{+2.5}_{-1.3}$ 	& & 20$\pm$3			& 1.0$^{+0.9}_{-0.6}$	& 9.1$^{+2.9}_{-1.3}$  			\\   
18 	& 1.407$\pm$0.004	& 8.96$\pm$0.16		& 2.3$\pm$0.4			& 12.9$\pm$0.9	& & \nodata 				& \nodata		\bigstrut[tb]				& \nodata				& & 16$\pm$3			& 0$^b$					& 18.0$\pm$1.1 			\\  
19 	& 1.516$\pm$0.005	& 12.8$\pm$0.2		& 2.5$\pm$0.5			& 12.0$\pm$1.0	& & \nodata 				& \nodata		\bigstrut[tb]				& \nodata				& & 22$\pm$8			& 0$^b$					& 13.1$^{+2.0}_{-0.9}$ 			\\  
20 	& 1.631$\pm$0.009	& 17.0$\pm$0.4		& 4.6$^{+2.6}_{-1.5}$ 	& 9.4$\pm$1.5	& & \nodata 				& \nodata		\bigstrut[tb]				& \nodata				& & 28$^{+45}_{-12}$ 	& 0$^b$					& 13.6$^{+4.8}_{-1.2}$ 			\\  
21 	& 1.729$\pm$0.015	& 25.3$\pm$1.2		& 3.5$\pm$1.6			& 11.3$\pm$1.7	& & \nodata 				& \nodata		\bigstrut[tb]				& \nodata				& & \nodata				& \nodata 				& \nodata 				 
\enddata 
\tablenotetext{a}{Number of frequency-selected group.}
\tablenotetext{b}{Value was fixed.}
\end{deluxetable*}
\end{center}

As a result of the increased signal-to-noise we were able to detect additional features in the power spectra, especially for the power spectra at low \sa. In Figure \ref{fig:pds} we show three representative power spectra, in which the $\sim$1 Hz QPO has frequencies of $\sim$0.5 Hz, $\sim$1.4 Hz, and $\sim$6.6 Hz (top, middle, and bottom panel, respectively). Fits to these power spectra are shown as well, including the individual components (a power law and up to five Lorentzians). As can be seen, in all cases the $\sim$1 Hz QPO is accompanied by several broad components and a power-law noise component. Based on a comparison with other NS-LMXBs (see Section \ref{sec:atoll}) and following the nomenclature used in prior works, we identify the following components in the power spectrum in the top panel of Figure \ref{fig:pds}: a break ($L_{\rm b}$) , the $\sim$1 Hz QPO, which we identify as a LF-QPO ($L_{\rm LF}$), a `hump' ($L_{\rm h}$), and two broad features that are commonly associated with the lower and upper kHz QPOs ($L_{\rm l}$ and $L_{\rm u}$). Given the quality of the power spectra, the lower and upper kHz QPOs are typically not well constrained and they will not be discussed in detail. The fit parameters of the $\sim$1 Hz QPO ($L_{\rm LF}$) and the two low-frequency noise components ($L_{\rm b}$ and $L_{\rm h}$) are given in Table \ref{tab:pds}. The power-law component gradually steepened from $\alpha\approx1.5$ at low QPO frequencies to $\alpha\approx2.5$ at high QPO frequencies, while it decreased in strength from $\sim$7\%--8\% to $\sim$2\%--3\%.

In Figure \ref{fig:sz_meta} we once again show the evolution of the $\sim$1 Hz QPO properties as a function of \sa, this time using the measurements from the frequency-selected groups. As can be seen from the top panel, combining groups based on frequency greatly reduces the scatter at low \sa\ compared to Figure \ref{fig:sz}, although some scatter is still present. Also, for the same reason no data points are present any longer below \sa\,=\,1.09. As before, the QPO frequency shows a strong increase with \sa. The $Q$-value decreases slowly up to \sa\,$\approx$\,1.4--1.6 and appears to increase after that. Although the fractional rms amplitude of the QPO still shows considerable scatter at low \sa\ values, overall there is an increasing trend up to \sa\,$\approx$\,1.4. The rms appears to drop again toward higher \sa\ values.

The $L_{\rm h}$ noise component is detected in all frequency-selected power spectra, except for the last one.  The $L_{\rm b}$ noise component is consistently detected up to QPO frequencies of $\sim$1.2 Hz. At higher QPO frequencies the $L_{\rm h}$ component becomes broader and as a result the $L_{\rm b}$ component becomes more difficult to detect. It is only detected twice more, when the  $L_{\rm h}$ component is narrower.

\begin{figure*}[t] 
\centerline{\includegraphics[height=7.cm]{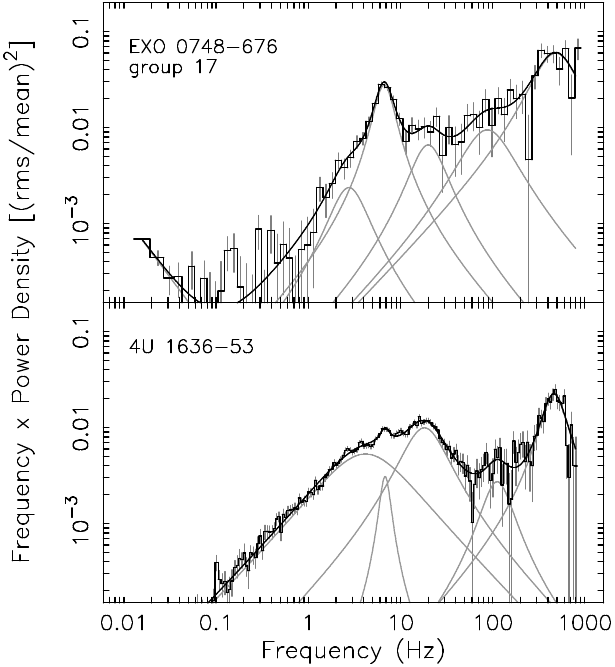}
\includegraphics[height=7.cm]{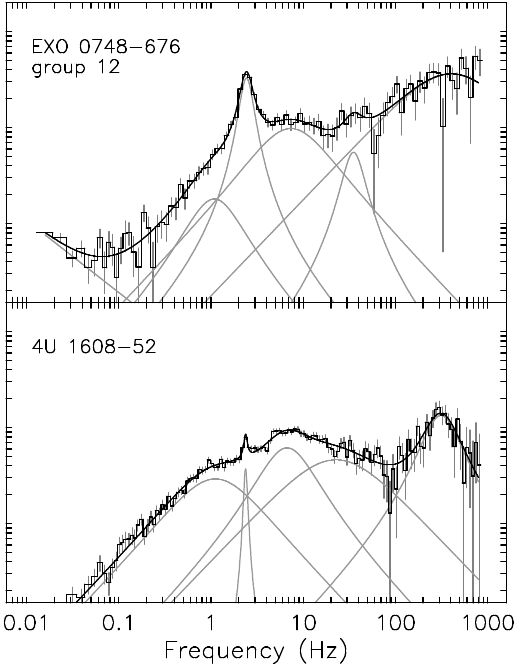}
\includegraphics[height=7.cm]{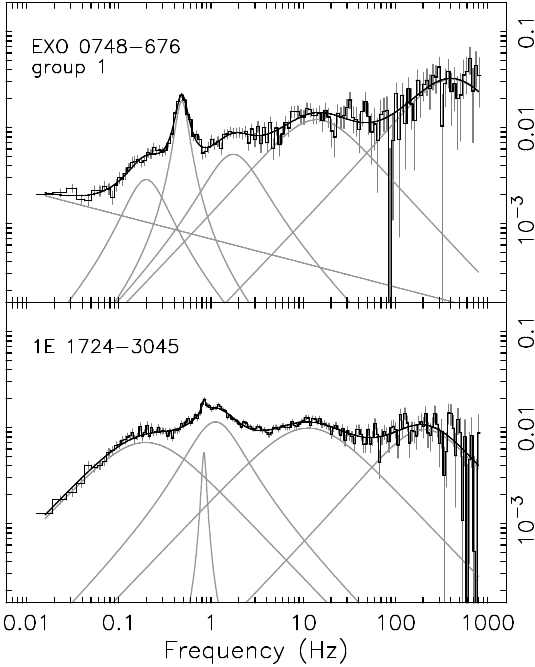}}
\caption{A comparison of power spectra of  \exo\ with those of three atoll sources. The following \xte\ ObsIDs were used for the atoll sources:  60032-05-21-00(0) for 4U 1636--53, 30062-02-02-00(0) for 4U 1608--52, and 10090-01-01-000 for 1E 1724--3045. } 
\vspace{0.2cm}
\label{fig:comparison}
\end{figure*}

\subsection{Comparison with other Atoll Sources}\label{sec:atoll}

Since \exo\ shows atoll-like patterns in its CD and HID, we investigated whether the $\sim$1 Hz QPO in \exo\ is possibly related to the LF-QPOs seen in the non-dipping/eclipsing atoll sources. Due to their relatively low rms amplitudes ($\sim$few percent) detections of LF-QPOs in atoll sources are not very common (especially compared to the LF-QPOs in the more luminous Z sources). We first searched the literature for \xte\ observations of non-dipping/eclipsing atoll sources with LF-QPOs that have frequency ranges (partially) overlapping that of the $\sim$1 Hz QPO in \exo. Several examples were found: 4U 1636--53 \citep{alvame2008a}, 4U 1608--52 \citep{vavame2003}, 1E 1724--3045 \citep{alvame2008b}, and 4U 1820--30 \citep{alvame2005}. In addition we also analyzed observations of 4U 1812--12 and found LF-QPOs around 0.3 Hz. In Figure \ref{fig:comparison} we compare the power spectra of the first three of these atoll sources with three power spectra of \exo. 

For similar QPO frequencies we find that the power spectra of \exo\ and the three atoll sources are to a large extent similarly shaped, with the accompanying noise and (broad) QPO components having similar frequencies as well. Most importantly, like the $\sim$1 Hz QPO in \exo, the LF-QPO component in the three atoll sources has a frequency between that of the \lb\ and \lh\ noise components (although in 1E 1724--3045 it is closer to the \lh\ component than in the other sources). However, there are several differences as well. The $\sim$1 Hz QPO in \exo\ is considerably stronger than the LF-QPOs detected in the atoll sources (in the same frequency range). As can be seen in Table \ref{tab:pds}, the $\sim$1 Hz QPO in \exo\ has a fractional rms amplitude in the range $\sim$9\%--15\%. The atoll source LF-QPOs, on the other hand, have fractional rms amplitudes of $\sim$2\%--5\%. On average the $Q$-values of the atoll sources LF-QPOs tend to be higher than those of the $\sim$1 Hz QPO in \exo; this is especially clear for the examples shown in Figure \ref{fig:comparison}. In \exo\ the $Q$-values ranged from $\sim$1.8 to  $\sim$4.6, while in the atoll source values of $\sim$1.6--8.8 are found (see above references). While grouping of observations may have led to some broadening of the QPO in \exo, similar $Q$-values were already reported for QPO detections in individual observations by \citet{hojowi1999}.

There are some differences in the behavior of the other power spectral components as well. The power spectra of \exo\  contain a clear power-law noise component, which is absent in the atoll source power spectra. Although some atoll sources do show such a power-law component in their power spectra, it typically does not appear until the source is closer to the soft state \citep[see, e.g.,][]{vavame2003}. We note that strong power-law components are present in some of the other dipping and/or eclipsing sources as well \citep{jovawi1999,jovaho2000}, suggesting its presence is related to viewing angle. Another difference pertains to the two broad high-frequency components (which we tentatively identified as the lower and upper kHz QPOs); these tend to be stronger in \exo\ than in the other atoll sources, as can be seen from Figure \ref{fig:comparison}.

\begin{figure*}[t] 
\centerline{\includegraphics[width=12.5cm]{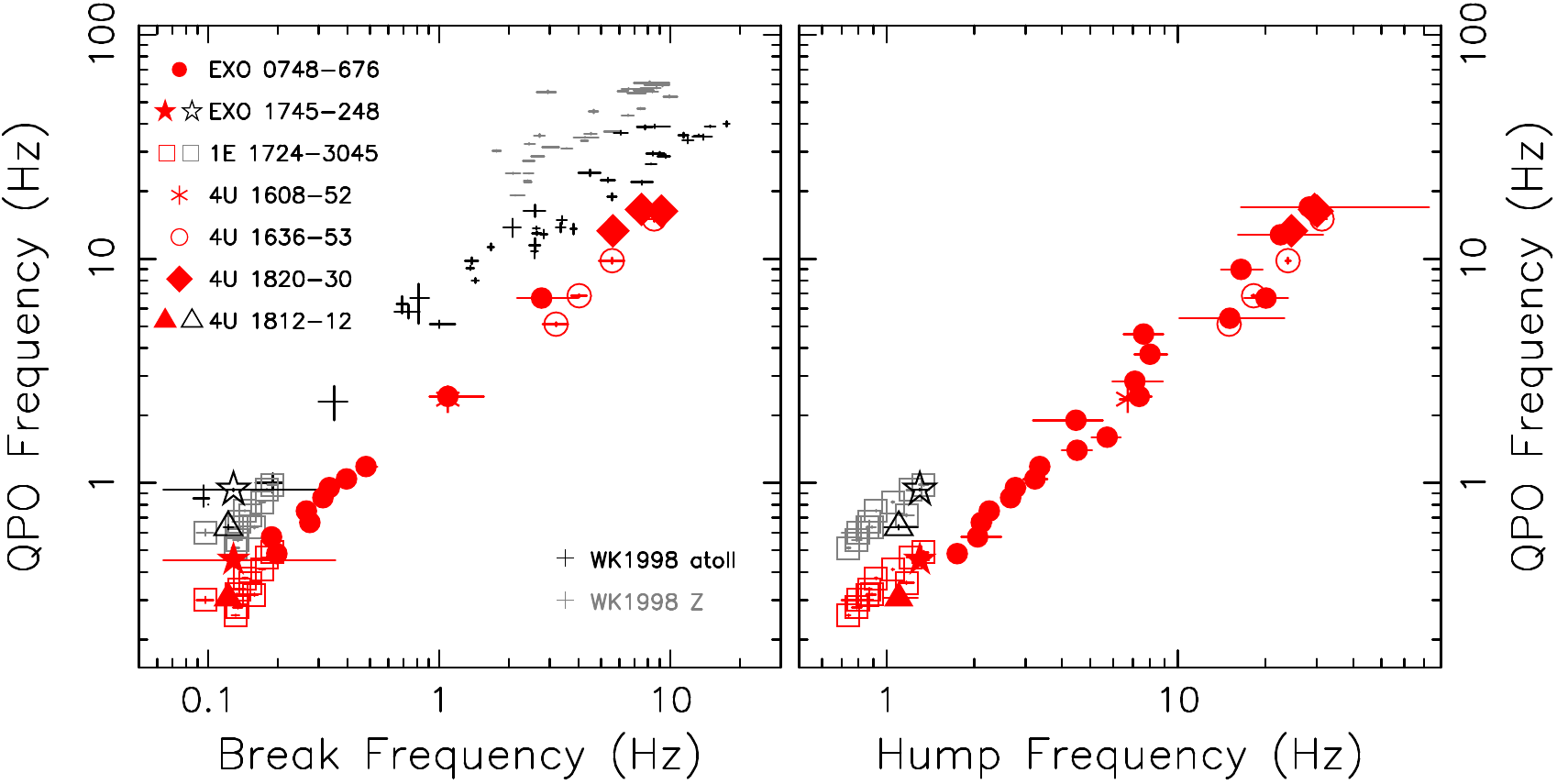}}
\caption{QPO frequency vs. break frequency (left panel) and vs.\ hump frequency (right panel). See left panel for the meaning of the various symbols. For reference we also show the original data from \citet{wiva1999} for atoll and Z sources. For 1E 1724--3045 (gray open squares) we also show the QPO frequencies divided by a factor of 2 (red open squares). Data were taken from \citet{alvame2008a} for 4U 1636--53,  \citet{alvame2005} for 4U 1820--30, and  \citet{alvame2008b} for \ter. For 4U 1608--52 we analyzed \xte\ observations  30062-02-02-00(0), for 4U 1812--12 observations 30701-13-01-[00:02], and for EXO 1745--248 observation 50054-06-04-03. } 
\label{fig:wk}
\end{figure*}

As the $\sim$1 Hz QPO in \exo\ and LF-QPOs in the atoll sources appear to have similar frequencies relative to those of the two main (\lb\ and \lh) noise components, it is interesting to explore whether the QPOs also follow similar relations in QPO frequency vs.\ noise frequency diagrams. Such studies have been performed for various power-spectral components in NS-LMXBs and BH-LMXBs \citep[see, e.g.,][]{psbeva1999,wiva1999,bepsva2002}. In Figure \ref{fig:wk} we plot the frequency of the QPOs in \exo\ and several atoll sources as a function of the frequency of the break component \lb\ (left panel) and as a function of the frequency of the hump component \lh\ (right panel). In addition to the atoll sources mentioned earlier, we also include data from the dipping NS-LMXB EXO 1745--248 (stars), which also shows a $\sim$1 Hz QPO \citep{ho2012}. For this source the break frequency was poorly constrained. Finally, for reference we show in the left panel of Figure \ref{fig:wk}  the original data of \citet{wiva1999} for the Z and atoll sources. We note, however, that for the power spectra of the atoll sources  \citet{wiva1999} mostly used the frequency of \lh\ for that of the QPO.  As a result the QPO values that we plot for \exo\ and the atoll sources fall well below the relation found by \citet{wiva1999}. 

As can be seen from both panels in Figure \ref{fig:wk}, \exo\ traces out relations that are largely consistent with those of the other atoll sources. We note that \ter\ (gray squares) presents somewhat of an outlier in both panels. The QPO frequency for this source lies a factor of $\sim$2 above that of the relation traced out by the other sources, suggesting that the QPO in \ter\ might be the second harmonic, with the fundamental not being present. In 4U 1812--12 (triangles) and EXO 1745--248 (stars) we detect the fundamental as well as the second harmonic. Both are shown in Figure \ref{fig:wk}, with the fundamentals falling on the main relation and the second harmonics coinciding with the cluster of \ter\ data points. This appears to confirm the harmonic nature of the \ter\ QPO. For this reason we also show the QPO frequencies of \ter\ divided by a factor of 2 in Figure \ref{fig:wk} (red squares); as can be seen, in this way the \ter\ data line up well with the main relation.

\section{Discussion}\label{sec:disc}

We have presented a study of the $\sim$1 Hz QPO in the dipping and eclipsing NS-LMXB \exo. The main goal of this work was to understand how this QPO evolves between the various spectral states and how its properties compare to those of the LF-QPOs in other NS-LMXBs.

 In its HID and CD \exo\ traces out tracks that strongly resemble those of the atoll source NS-LMXBs, with hard (extreme island), transitional (island), and soft (banana) states. We found that the $\sim$1 Hz QPO in \exo\ is only observed in the hard and intermediate states and that its frequency strongly increases as the spectrum softens. After combining power spectra based on QPO frequency, a relatively tight relation with the track parameter \sa\ is observed (see Figure \ref{fig:sz_meta}), with the frequency increasing from  $\sim$0.5 Hz at \sa\,$\approx$\,1.2 to $\sim$25 Hz at \sa\,$\approx$\,1.8. This increase in frequency  from the hard state into the intermediate state, as well as the frequency range covered, is similar to what is observed for the LF-QPOs in other atoll sources in these states \citep{vavame2003,alvame2005,alvame2008a,alvame2008b}.  

The grouping of power spectra from individual ObsIDs (first based on time and later also based on QPO frequency) resulted in high-quality power spectra that revealed several peaked noise components in addition to the QPO and power-law noise component already reported in the $\sim$1 Hz QPO discovery papers. A comparison of the QPO and noise components in the power spectra of \exo\ with those in atoll source power spectra revealed strong overall similarities. Most importantly, like the LF-QPO in atoll sources, the $\sim$1 Hz QPO in \exo\ ($L_{\rm LF}$) is located between the break (\lb) and hump (\lh) noise components (Figure \ref{fig:comparison}) and its frequency evolves in a similar way with respect to that of the two noise components (Figure \ref{fig:wk}).

These results strongly suggest that the frequencies of the $\sim$1 Hz QPO in \exo\ and the LF-QPOs in (other) atoll sources are set by the same mechanism. From past work \citep{hojowi1999,jovawi1999,jovaho2000} it is clear that $\sim$1 Hz QPOs in \exo\ and other dipping/eclipsing NS-LMXBs have a geometric nature, in the sense that parts of the accretion flow continuously move (rotate) in and out of our line of sight to the neutron star, thus modulating the emission we see from the neutron star surface and/or boundary layer. \citet{ho2012} suggested that a misaligned precessing inner flow could provide such a continuously changing geometry. 

Since the non-dipping/eclipsing atoll sources have lower inclination angles, and there is no evidence for line of sight modulations, such a precessing inner disk would have to modulate the X-ray emission in a different manner in those sources. Modulation of the X-rays through relativistic beaming and light-bending effects, as suggested for LF-QPOs in BH-LMXBs \citep{schomi2006}, could offer an explanation for the non-dipping/eclipsing sources. We note that these effects are likely also at work in dipping/eclipsing sources. However, Figure 3 in \citet{schomi2006} suggests that the resulting modulations are out of phase with the modulations caused by the line of sight interceptions. This may in fact reduce the overall strength of the $\sim$1 Hz QPO, although it is still observed to be significantly stronger than the LF-QPOs in non-dipping/eclipsing atoll sources. The differences in modulation mechanism between the two types of QPOs may also explain the (on average) higher rms amplitudes and lower quality factors of the $\sim$1 Hz QPOs in comparison to the LF-QPOs in the non-dipping/eclipsing atoll sources, although detailed simulations are necessary to verify this. In addition to resulting in different modulation mechanisms for the QPOs, we speculate that the higher inclination of \exo\ may also be the cause of the more prominent power-law noise component as well as differences in the relative strengths of the broad peaked noise components, although the mechanisms involved are unclear. 

As mentioned in Section \ref{sec:intro}, there is increasing evidence for a geometric origin of the (type-C) LF-QPOs in BH-LMXBs \citep{heutkl2014,inva2015,mocahe2015}, with Lense--Thirring precession \citep{stvi1998} currently being the most promising mechanism \citep{indofr2009}. While we argued above that the $\sim$1 Hz QPO in \exo\ and the LF-QPOs in other atoll sources have a geo\-metric nature as well, it is not clear to what extent their frequencies are set, or dominated, by Lense--Thirring precession. Classical precession (due to neutron-star oblateness) and magnetic precession (due to magnetic torques) may significantly lower the precession frequency \citep{la1999,shla2002}, while radiation feedback from the neutron star could significantly increase the precession frequency \citep{mi1999b}. The latter mechanism may explain why, for the same break (\lb) frequency, the LF-QPOs in the more luminous Z sources (the horizontal-branch oscillations) have higher frequencies (by a factor of $\sim$5, see Figure \ref{fig:wk}) than the $\sim$1 Hz QPO and  LF-QPOs in \exo\ and other atoll sources. In fact, \citet{alinva2012} already pointed out that the frequencies of the horizontal-branch oscillations in the Z source and 11 Hz pulsar IGR J17480--2446 were too high to be explained by pure Lense--Thirring precession, suggesting that radiation feedback can indeed significantly alter (or even dominate) the precession frequency in neutron star systems. However, we stress that the conclusions we have drawn about the geometric nature of the $\sim$1 Hz QPO and LF-QPOs in atoll  sources do not necessarily apply to the horizontal-branch oscillations in the Z sources, since these conclusions were based on the similarities between the power spectra of \exo\ and the non-dipping/eclipsing atoll sources.  The horizontal-branch oscillations follow a very different relation in the left panel of Figure  \ref{fig:wk} (although with a similar slope) and it is therefore not clear whether they are the result of a similar mechanism as the LF-QPOs in atoll sources.

In none of the seven NS-LMXBs that have shown \hbox{$\sim$1 Hz} QPOs are the QPOs detected in the soft state \citep{ho2012}. Combined with the fact that LF-QPOs in non-dipping/eclipsing atoll sources are also rare (or absent) in the soft state, this suggests that the precessing part of the accretion flow in atoll NS-LMXBs becomes significantly smaller toward the soft state and/or that  the misalignment angle between the inner flow and neutron-star spin axis decreases.  Both changes would  reduce the effects of line of sight interceptions in the dipping/eclipsing systems and of the relativistic beaming and light-bending effects in non-dipping/eclipsing sources.
In the case of the $\sim$1 Hz QPO a reduction in the geometrical thickness and/or the optical thickness of the inner flow may also affect the efficiency of removing photons from the neutron star out of our line of sight and therefore reduce the strength of the QPO. Finally, the frequency increase of the $\sim$1 Hz QPOs in \exo\ and the LF-QPOs in the atoll sources indicates that the transition radius between the (standard) accretion disk  and the precessing inner accretion flow decreases as the spectrum softens, similar to what has been concluded from the type-C QPOs in BH-LMXBs \citep{indo2011b}.

\section{Conclusions}\label{sec:sum}

We have studied the $\sim$1 Hz QPO in the dipping and eclipsing NS-LMXB \exo\ and compared it to the LF-QPOs in non-dipping/eclipsing atoll NS-LMXBs. We find strong similarities between the two types of QPOs and conclude that their frequencies are likely set by the same mechanism. A misaligned precessing inner accretion flow is suggested as the common frequency mechanism; the resulting modulations are geometric in nature, but the way in which the modulations occur depends on viewing angle. In the (low-inclination) non-dipping/eclipsing sources the emission from the precessing flow could be modulated through relativistic beaming and light-bending effects, while in the  (higher-inclination) dipping/eclipsing systems we get an additional (and dominating) modulation of the X-rays from the neutron star through line of sight interceptions by the precessing inner flow.



\acknowledgments

J.H.\ acknowledges financial support from NASA grant NNX12AE14G, provided through the Astrophysics Data Analysis Program. This research has made use of data obtained from the High Energy Astrophysics Science Archive Research Center (HEASARC), provided by NASA's Goddard Space Flight Center.


\end{document}